\documentclass[aps,twocolumn,amsmath,amssymb,showpacs,preprintnumbers]{revtex4}
\usepackage{graphicx,psfrag,color,ulem}
\usepackage{amssymb,latexsym,mathrsfs,amsmath}
\begin{document}
\def\clr{\textcolor{red}}
\def\be{\begin{equation}}
\def\ee{\end{equation}}
\def\bea{\begin{eqnarray}}
\def\eea{\end{eqnarray}}
\def\et{\rho}
\def\n{\nonumber}
\def\c{\mathscr}
\def\eps{\delta}
\def\bra #1{\langle #1|}
\def\ket #1{|#1\rangle}
\def\et{\rho}
\def\fr{\frac}
\newcommand{\ra}{\rangle}
\newcommand{\la}{\langle}

\title{Driven $k$-mers: Correlations in space and time}
\author{Shamik Gupta,$^{1,2,3}$ Mustansir Barma,$^1$ Urna Basu,$^4$ and P. K. Mohanty$^4$}
\affiliation{
{\small $^1$Department of Theoretical Physics, Tata Institute of
Fundamental Research, Homi Bhabha Road, Mumbai 400005, India}\\
{\small $^2$Physics of Complex Systems, Weizmann Institute of Science, Rehovot 76100, Israel}\\
{\small $^3$Laboratoire de Physique de l'\'{E}cole Normale Sup\'{e}rieure de Lyon, Universit\'{e}
de Lyon, CNRS, 46 All\'{e}e d'Italie, 69364 Lyon c\'{e}dex 07, France} \\
{\small $^4$Theoretical Condensed Matter Physics Division, Saha Institute of Nuclear Physics, Kolkata 700064, India}}
\preprint{TIFR/TH/11-27}
\begin{abstract}
Steady-state properties of  hard objects with exclusion interaction and
a driven motion along a one-dimensional periodic lattice 
are investigated. The process is a generalization of the asymmetric simple exclusion process (ASEP) to particles of length $k$, and 
is called the $k$-ASEP. Here, we analyze both static and dynamic properties of the $k$-ASEP. Density correlations are found to display 
interesting features, such as pronounced oscillations in both space and time, as a consequence of the extended length of the particles. At long times, the density autocorrelation decays exponentially in time, except at a special $k$-dependent density when it decays as a power law. 
In the limit of large $k$ at a finite density of occupied sites, the appropriately scaled system reduces to a nonequilibrium generalization of the Tonks gas 
describing the motion of hard rods along a continuous line. This allows us
to obtain in a simple way the known two-particle distribution for the
Tonks gas. For large but finite $k$, we also obtain the leading-order
correction to the Tonks result. 
\end{abstract}
\date{\today}
\maketitle
\section{Introduction}
The asymmetric simple exclusion process (ASEP), 
a paradigmatic model of nonequilibrium statistical mechanics, involves
hard-core particles
 undergoing biased diffusion on a lattice in the presence of an external drive
 \cite{Liggett:1985,Schutz:2001,Mallick:2006,Zia:2011,Zia:20111}. The
 generalization of the ASEP to an exclusion process of hard-core extended objects ($k$-mers, each of which occupies $k$ consecutive sites) is referred to as the $k$-ASEP. It was first
introduced to model protein synthesis inside living cells \cite{Macdonald:1968, Macdonald:1969}. During the synthesis, 
ribosomes move from codon to codon along messenger RNA, read off genetic
information, and generate the protein stepwise. Modelling the codons by lattice sites and the ribosomes by $k$-mers, we recover the $k$-ASEP. The spatial extent of the $k$-mers takes 
care of the blocking of several codons by a single ribosome; steric hindrance, which prevents overlap of ribosomes, is modelled by the 
exclusion constraint. 
 
Earlier studies of the $k$-ASEP in one dimension involved analyzing the
steady-state density profile in an open system \cite{Macdonald:1968, Macdonald:1969}, 
the time-dependent conditional probabilities of finding the $k$-mers on specific sites at a given time \cite{Sasamoto:1998}, 
the dynamical exponent \cite{Alcaraz:1999}, the phase diagram of the system with open boundaries \cite{Lakatos:2003, Shaw:2003, Shaw:2004, Shaw:20042}, 
the hydrodynamic limit governing the evolution of the density
\cite{Schonherr:2004}, and the effects of defect locations on the
lattice on steady-state 
properties \cite{Zia:2007,Dongthesis}. Some aspects of the $k=2$ case of the $k$-ASEP were studied earlier in the context of a model of driven, reconstituting 
dimers \cite{Barma:2007}.  

Here, we are concerned with the $k$-ASEP on a one-dimensional ($1$D) periodic lattice. At long times, the process settles into a nonequilibrium steady state in which
 all configurations with a given number of $k$-mers have equal weights \cite{Shaw:2003}. In this work, our focus is on correlation functions, both static and dynamic.

We compute static correlations in two different ways: (i) by counting the number of relevant configurations, and (ii) by mapping the $k$-ASEP to a zero-range process (ZRP) \cite{Evans:2005} and then by employing a matrix product formalism \cite{Basu:2010}. Dynamic correlations in the $k$-ASEP are derived by mapping the $k$-ASEP to an equivalent ASEP with a smaller number of sites and by using the 
known dynamic properties of the latter \cite{Barma:2007}. We show that density correlations exhibit pronounced oscillations in both space and time as a consequence of the extended length of the $k$-mers. 

One may also consider the $k$-ASEP in the continuum limit, i.e., in the joint limit of large $k$ and vanishing lattice spacing, $\eps \rightarrow 0$, while
  keeping the product $a=k\eps$ and the density of occupied sites fixed and finite. Such a limit was considered previously to obtain the hydrodynamic behavior of the continuum system \cite{Schonherr:2005}. In this limit, the model describes hard 
rods of finite length $a$, undergoing biased diffusion along a continuous line in the presence of an external drive. In the case of unbiased motion, this continuum model was 
studied earlier by Tonks \cite{Tonks:1936}. This so-called Tonks gas has an equilibrium steady state in which quantities of physical interest, e.g., 
the equation of state and the two-particle distribution function, have been worked out exactly \cite{Tonks:1936,Salsburg:1953}. The continuum limit of 
the $k$-ASEP is a nonequilibrium generalization, and may be called the driven Tonks gas. 

The known two-particle distribution function of the Tonks gas is
recovered straightforwardly by taking the continuum limit of the
$k$-ASEP, on noting that the steady-state measure of configurations is
the same, whether or not the system is driven. We also obtain the
leading-order correction to the Tonks result
when $k$ is large but finite. The Tonks result, after including the
leading-order correction in $1/k$, turns out to be a good approximation
to the $k$-ASEP equal-time spatial correlation even for not too large values of
$k$ (for example, $k=13$ when the density of occupied sites is $0.75$).
 
The paper is organized as follows. In Sec.
\ref{themodelrelationtoDDRDsteadystate}, we define the $k$-ASEP and
discuss its steady-state measure. In the following section, we derive
closed-form expressions for the steady-state equal-time engine-engine
and density-density correlations. The former function describes
correlation between the right end (the ``engine") of one $k$-mer with
that of another at the same time instant. Considering the continuum
limit of the $k$-ASEP, we derive the known two-particle
distribution function for the Tonks gas. We also derive the leading
finite-$k$ correction to the Tonks result. In Sec. 
\ref{steadystatedynamics}, we address the steady-state dynamics by computing the $k$-mer current and kinematic wave velocity associated with transport of density fluctuations. We then discuss a mapping of the $k$-ASEP to an equivalent ASEP, and the behavior of the $k$-ASEP temporal density-density correlation, whose scaling properties are derived by utilizing the mapping. In Appendix \ref{ZRPmapping}, we discuss a different method to obtain static correlations in the $k$-ASEP through a mapping to an equivalent ZRP.  
\section{The $k$-ASEP}
\label{themodelrelationtoDDRDsteadystate}
\subsection{Definition}
\label{themodel}
We consider a number, $N$, of $k$-mers that are subject to hard-core
exclusion and are distributed on a $1$D periodic lattice of $L$ sites. 
The sites are labeled by the index $i=1,2,\ldots,L$. Each $k$-mer occupies $k$ consecutive lattice sites. The $k$-mers are hard 
objects that cannot break into smaller fragments. The density of occupied sites in the system is given by
\be
\rho \equiv \frac{Nk}{L}.
\label{rhodefn}
\ee
We specify the location of a $k$-mer on the lattice by the site index of
its rightmost end, and we call this end the ``engine" of the $k$-mer. 
We denote the occupation of the $i$th site by $n_i=1$ or $0$, according
to whether the site is occupied or vacant, respectively. A $k$-mer is then represented 
by a string of $k$ consecutive $1$'s and a configuration of the model by an $L$-bit binary string composed of $0$'s and $1$'s. 

\begin{figure}[h]
\begin{center}
\includegraphics[width=80mm]{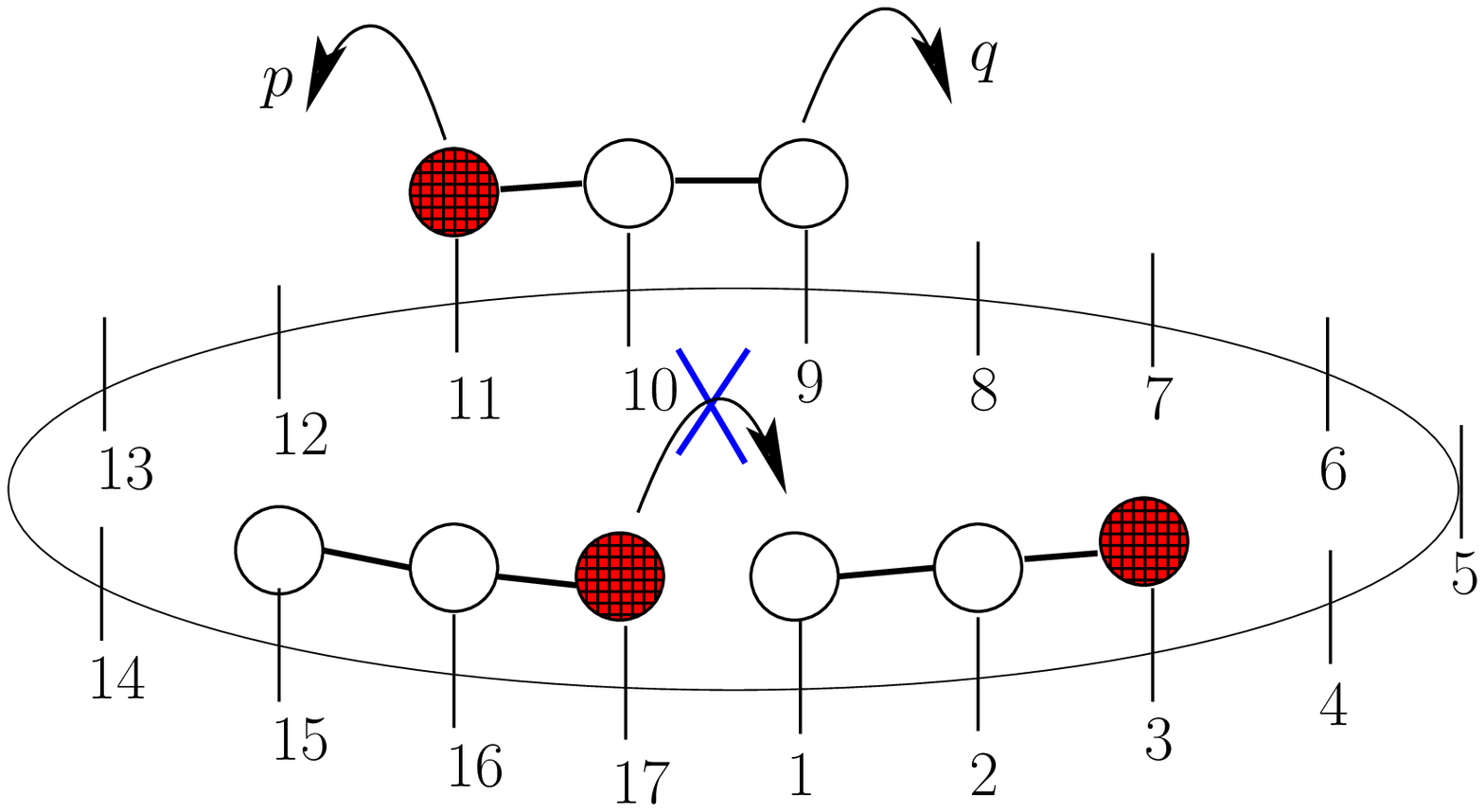}
\end{center}
\caption{(Color online) The $k$-ASEP on a ring, showing the allowed and disallowed dynamical moves of trimers ($k=3$). Here, $k$-mers are represented as $k$ connected circles. The rightmost end of a $k$-mer (the ``engine") is indicated by a filled red circle.}
\label{kmerdefinitionfig}
\end{figure}

The system of $k$-mers evolves according to a stochastic Markovian dynamics: in a small time $dt$, a $k$-mer advances forward (respectively, backward) by one lattice site with probability $pdt$ (respectively, $qdt$), provided the site is unoccupied. The dynamics conserves the total number of $k$-mers in the system. The elementary 
dynamical moves may be represented as
\be
(111\ldots1)0\overset{pdt}{\underset{qdt}{\rightleftarrows}}0(111\ldots1),
\label{evolutionrules}
\ee
where the $k$-mer has been represented by a string of $k$ consecutive $1$'s enclosed by brackets. Figure \ref{kmerdefinitionfig} illustrates the allowed and disallowed moves of trimers ($k=3$). 

For $p\ne q$, the $k$-mers move preferentially in one direction along the lattice, and the system at long times reaches a
 nonequilibrium steady state with a steady current of $k$-mers. 

When $p = q$, the $k$-mers diffuse symmetrically to the left and to the right. In this case, the model is a generalization of 
the symmetric simple exclusion process (SEP) of hard-core particles to $k$-mers, and may be referred to as the $k$-SEP. 
At long times, the $k$-SEP settles into an equilibrium steady state.   
\subsection{Relation to a model of diffusing, reconstituting dimers}
\label{relationtoDDRD}
Earlier studies of a variant of the $k$-ASEP dealt with dimers ($k=2$) that, in general, do not retain their identities and 
are allowed to reconstitute \cite{Barma:1997,Barma:2007}. Specifically, on a lattice of $L$ sites, single particles or 
monomers (denoted by $1$) and paired particles or dimers (denoted by $11$) are distributed with at most one particle per site. 
The dynamics in a small time $dt$ involves a dimer moving by one lattice site, either forward with probability $pdt$ or backward 
with probability $qdt$, without violating the hard-core constraint on site occupancies. The pairing of the dimers is impermanent, 
thereby allowing for reconstitution. For example, in the sequence of transitions, $11010 \rightarrow 01110 \rightarrow 01011$, 
the middle particle is paired with the particle to the left in the first transition and with the particle to the right 
in the second transition. 

Both in the symmetric ($p=q$) \cite{Barma:1997} and in the asymmetric ($p \ne q$) \cite{Barma:2007} case, the phase space of the system breaks up into an infinite number of dynamically disjoint sectors. A non-local construct, called the 
irreducible string (IS), uniquely labels the different sectors. The IS for a given configuration is constructed from the 
corresponding $L$-bit binary string by deleting recursively any pair of adjacent $1$'s until no further deletion is possible. The model 
exhibits dynamical diversity with quantities like the density autocorrelation showing strong sector-dependent behaviors that 
range from power laws to stretched exponentials. 

The problem of hard non-reconstituting dimers, i.e. the case $k=2$ of the $k$-mer system considered in this work, corresponds to a 
particular sector of the reconstituting dimer problem, namely, the one with the null IS $0000\ldots 000$. It is easy to check that this is 
the sector in which dimers do not reconstitute.  

\subsection{The steady state}
\label{steadystate}
In the steady state of the $k$-ASEP, every microscopic configuration $C$ with a given number of $k$-mers is equally likely, and 
hence, occurs with probability $1/\Omega$, where $\Omega$ is the total
number of configurations \cite{Shaw:2003}. In such a state, for every
transition away from $C$ to another configuration $C'$, it is simple to
construct a distinct and unique configuration $C''$ that 
evolves to $C$ at the same rate, thereby ensuring stationarity. This argument holds for periodic boundary conditions and fails in the case of open boundaries where the exact steady state is hard to obtain and is as yet unknown.  

Now, $\Omega$ is determined by counting the different possible ways of distributing a number, $N$, of $k$-mers over a 
lattice of $L$ sites with periodic boundaries. First consider {\it free} boundaries, in which case we have an open chain of $L$ sites on which the $k$-mers are placed. Then, the number of possible ways of distributing the $k$-mers is 
\be
\Omega^{\mathrm{free}}_{N,L}=\binom{L-Nk+N}{N}.
\label{omegafree}
\ee
For a periodic lattice, the number of ways is determined by first considering all configurations 
in which an arbitrary but fixed site remains occupied by an arbitrarily chosen but fixed engine. The number of such configurations 
is simply $\Omega^{\mathrm{free}}_{N-1,L-k}$. However, the engine could be chosen to be one of the $N$ identical engines available and 
may be placed over any of the available $L$ sites. Thus, the total number of distinct configurations is \cite{Shaw:2003}
\be
\Omega=\frac{L}{N}\Omega^{\mathrm{free}}_{N-1,L-k}=\frac{L}{N}\binom{L-Nk+N-1}{N-1}.
\label{omega}
\ee 

The steady-state probability that a randomly chosen site is occupied by an engine is obtained by considering all possible ways of distributing a number, $N-1$, of $k$-mers over a lattice of $L-k$ sites with free boundaries. 
Since all configurations of the $k$-ASEP are equally likely in the steady state, the desired probability is given by the ratio $\Omega^{\mathrm{free}}_{N-1,L-k}/\Omega=\rho/k$. 
Similarly, the probability that a randomly chosen site is vacant in the steady state may be shown to be equal to $1-\rho$. Next, we consider the joint probability that in the steady state, a randomly chosen site is occupied by an engine while the following site is vacant; this is given by the ratio $\Omega^{\mathrm{free}}_{N-1,L-(k+1)}/\Omega$. In the thermodynamic limit, i.e. in the limit $N \rightarrow \infty, L \rightarrow \infty$, while keeping 
$k$ and $\rho$ fixed and finite, this joint probability equals $\rho(1-\rho)/[\rho+k(1-\rho)]$.

We note that for the $k$-SEP, the condition of detailed balance implies that all configurations with a given number of $k$-mers have equal weights in its equilibrium steady state. 
It thus follows that the exclusion process with $k$-mers has the same steady state, irrespective of whether the $k$-mers have a biased or an unbiased motion. 
\section{Steady state statics}
\label{steadystatestatics}
\subsection{Engine-engine correlation}
Let $E_i$ denote the engine occupation variable for site $i$, taking
values $1$ or $0$ according to whether the site is occupied by an engine or not, 
respectively. The (unsubtracted) equal-time engine-engine correlation function $\mathcal{E}_k(r)$ is defined as
\be
\mathcal{E}_k(r)\equiv\langle E_{i}E_{i+r} \rangle,
\label{erdefinition}
\ee 
where the angular brackets denote averaging with respect to the steady state. 
Note that $\mathcal{E}_k(r)$ is identically zero for $r < k$. 

Now, $\mathcal{E}_k(r)$ for $r \ge k$ has non-zero contributions from all configurations in which there are two engines at 
the $i$-th and $(i+r)$-th sites, with the gap between the two engines containing $r-k$ sites that are occupied by any number, 
$m$, of $k$-mers between $0$ and the maximum number that may be placed
in the gap, while the left over $N-(m+2)$ number of $k$-mers is 
distributed over the remaining $L-(r-k+2k)$ sites. The maximum number of $k$-mers that may be placed over the gap of $r-k$ sites is given 
by $\lfloor (r-k)/k \rfloor$, where $\lfloor x \rfloor$ denotes the floor function that gives the largest integer not greater than $x$. 
Noting that in the steady state of the $k$-ASEP, all configurations have equal weights, we get 
\be
\mathcal{E}_k(r)=\sum_{m=0}^{\lfloor{(r-k)/k}\rfloor}\frac{\Omega^{\mathrm{free}}_{m,r-k}\Omega^{\mathrm{free}}_{N-m-2,L-r-k}}{\Omega}.
\label{erexact}
\ee

In the thermodynamic limit, one can show by using 
Eq. (\ref{omegafree}) that, for arbitrary integers $m_1$ and $m_2$,
\be
\frac{\Omega^{\mathrm{free}}_{N-m_1,L-m_2}}{\Omega^{\mathrm{free}}_{N,L}} = \rho_0^{m_1}(1-\rho_0)^{m_2-m_1k},
\label{omegaapprox}
\ee   
where 
\be
\rho_0 \equiv \frac{\rho}{k(1-\rho)+\rho}. \label{eq:rho0}
\ee
Later, in Sec. \ref{steadystatedynamics}, by using a mapping of the $k$-ASEP to  the usual ASEP, we show that the quantity $\rho_0$ is in fact the particle density in the latter. Using Eq. (\ref{omegaapprox}), we find that in the thermodynamic limit, Eq (\ref{erexact}) reduces to 
\bea
&&\mathcal{E}_k(r) = \frac{\rho_0^2}{1+\rho_0(k-1)}\nonumber \\
&&\hspace{-0.5cm}\times\sum_{m=0}^{\lfloor{(r-k)/k}\rfloor}\binom{r-k-km+m}{m}\rho_0^m(1-\rho_0)^{r-k-km}; ~~r \ge k. \nonumber\\
\label{erthermodynamiclimit}
\eea

\begin{figure}
\includegraphics[width=90mm]{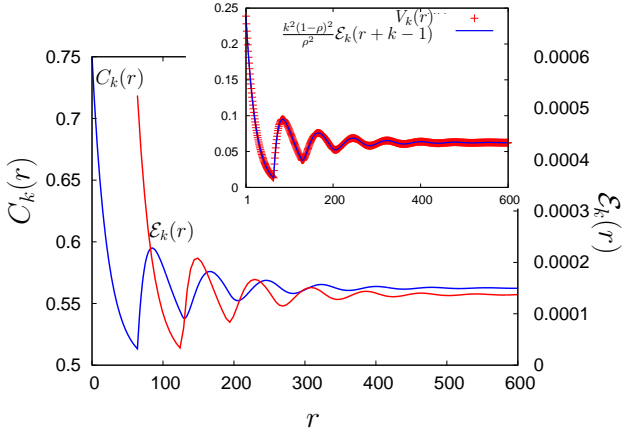}
\caption{(Color online) The $k$-ASEP density-density correlation $C_k(r)$ and engine-engine correlation 
$\mathcal{E}_k(r)$, evaluated numerically by using Eqs. (\ref{crexpression}) and (\ref{erthermodynamiclimit}) for $k=64$ and $\rho=0.75$. The inset shows that $\mathit{V}_k(r)$ and $[k^2(1-\rho)^2/\rho^2]\mathcal{E}_k(r+k-1)$ are equal, in agreement with Eq. (\ref{vrerrelation}). 
}
\label{ercrunscaled}
\end{figure}

For $k=2$, the sum in Eq. (\ref{erthermodynamiclimit}) can be evaluated exactly to 
obtain
$\mathcal{E}_k(r)=\frac{\rho_0^{2}}{(1+\rho_0)^{2}}[1-(-\rho_0)^{r-1}]$
\cite{Barma:2007}. For $k \ge 3 $, the function $\mathcal{E}_k(r)$ may
be numerically evaluated by using Eq. (\ref{erthermodynamiclimit}).
Figure \ref{ercrunscaled} shows the result for $k=64$. We see that the
engine-engine correlation exhibits damped oscillations in space, a
hallmark of systems with hard-core interactions between the constituents
\cite{Barrat:2003}, where purely entropic considerations apply. For
example, the fact that $\mathcal{E}_k(r)$ has its first minimum at
$r=2k-1$ and rises again at $r=2k$ may be understood to be due to the
possibility that when $r=2k$, there may be configurations with an
additional engine between the two engines that are occupying sites $i$ and $i+2k$. 

\subsection{Density-density correlation}
\label{densitydensityengineenginecorrelation}
The (unsubtracted) equal-time density-density correlation function $C_k(r)$ is defined as
\be
C_k(r)\equiv\langle n_{i}n_{i+r} \rangle.
\label{crdefinition}
\ee 
Here, $n_i$ is the occupation variable for site $i$, taking values $1$
or $0$, according to whether the site is 
occupied or is vacant, respectively. Evidently, $n_i$ can be expressed in terms of the engine occupation variable $E_i$ as
\be
n_i=\sum_{m=0}^{k-1}E_{i+m}.
\ee
Using the above equation, $C_k(r)$ may be expressed in terms of the equal-time engine-engine correlation as
\be
C_k(r)=k\mathcal{E}_k(r)+\sum\limits_{m=1}^{k-1}m\Big[\mathcal{E}_k(r+k-m)+\mathcal{E}_k(r-k+m)\Big],
\label{crexpression}
\ee
where it is understood that $\mathcal{E}_k(r)$ is zero for $r < k$.  

Alternatively, $C_k(r)$ can be computed in a straightforward way from the equal-time vacancy-vacancy correlation, defined as 
\be
\mathit{V}_k(r) \equiv \langle \overline{n}_i \overline{n}_{i+r} \rangle,
\label{vrdefinition}
\ee
where $\overline{n}_i=1-n_i$, so that 
\be
C_k(r)=\mathit{V}_k(r)+(2\rho-1).
\label{vrcrrelation}
\ee
Now, $\mathit{V}_k(r)$ for $r \ge 1$ has non-zero contributions from all configurations in 
which both the $i$-th and $(i+r)$-th sites are vacant, with the gap between the two containing $r-1$ sites 
that are occupied by any number, $m$, of $k$-mers between $0$ and the maximum number that may be placed in the gap, 
while the left over $N-m$ number of $k$-mers is distributed over the remaining $L-(r-1+2)$ sites. We get
\be
\mathit{V}_k(r)=\sum_{m=0}^{\lfloor{(r-1)/k}\rfloor}\frac{\Omega^{\mathrm{free}}_{m,r-1}\Omega^{\mathrm{free}}_{N-m,L-r-1}}{\Omega}.
\label{vrexact}
\ee

In the thermodynamic limit, we find that  
\bea
&&\mathit{V}_k(r) = \frac{(1-\rho_0)^2}{1+\rho_0(k-1)} \nonumber \\
&&\hspace{-0.5cm}\times\sum_{m=0}^{\lfloor{(r-1)/k}\rfloor}\binom{r-1-km+m}{m}\rho_0^m(1-\rho_0)^{r-1-km};~~r \ge 1. \nonumber \\
\label{vrthermodynamiclimit}
\eea
It follows from the definition that $\mathit{V}_k(0)=1-\rho$. 
On comparing Eq. (\ref{vrthermodynamiclimit}) with Eq.
(\ref{erthermodynamiclimit}), and noting that
$(1-\rho_0)^2/\rho^2_0=k^2(1-\rho)^2/\rho^2$, we get
\be
\mathit{V}_k(r)=\frac{k^2(1-\rho)^2}{\rho^2}\mathcal{E}_k(r+k-1); ~~r \ge 1.
\label{vrerrelation}
\ee
Using Eq. (\ref{vrcrrelation}), we get
\be
C_k(r)=\frac{k^2(1-\rho)^2}{\rho^2}\mathcal{E}_k(r+k-1)+(2\rho-1); ~~r \ge 1.
\label{crcompact}
\ee
For $r=0$, we have $C(0)=\rho$.

Figure \ref{ercrunscaled} shows $C_k(r)$,  computed  using Eqs. \eqref{crexpression} and \eqref{erthermodynamiclimit}, 
for $k=64$ and $\rho=0.75$. The oscillations in $C_k(r)$ may be related to those in $\mathcal{E}_k(r)$ by using Eq. (\ref{crcompact}). For instance, since $\mathcal{E}_k(r)$ has its first minimum at $r=2k-1$, it follows that the first minimum of $C_k(r)$ occurs at $r=k$. The inset of Fig. \ref{ercrunscaled} shows that  
$\mathit{V}_k(r)$ and $[k^2(1-\rho)^2/\rho^2]\mathcal{E}_k(r+k-1)$ for
$r \ge 1$ are equal in accordance with
Eq. (\ref{vrerrelation}).

Later, in Appendix \ref{ZRPmapping}, we discuss an alternative method to obtain static correlations in the $k$-ASEP through a mapping to a zero-range process and employing a matrix product formalism \cite{Basu:2010}.
\subsection{Continuum limit: Driven Tonks gas}
\label{driventonksgas}
In a suitable continuum limit of the $k$-ASEP, discussed below, the model reduces to one of hard rods, which have exclusion interaction, and 
 which are undergoing driven, diffusive motion along a continuous line. In the absence of drive, this continuum model was 
studied by Tonks as a $1$D interacting system with an equilibrium steady state in which thermodynamic properties like the 
equation of state can be worked out exactly \cite{Tonks:1936}. When the motion is driven, the continuum limit of the $k$-ASEP 
becomes the driven Tonks gas. 

\begin{figure}
\begin{center}
\includegraphics[width=80mm]{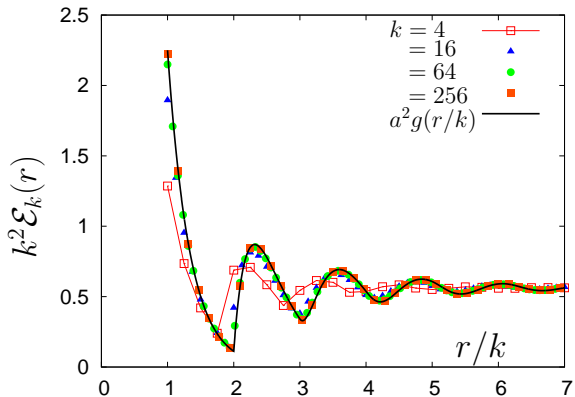}
\caption{(Color online) Scaling approach of the engine-engine
correlation $\mathcal{E}_k(r)$ to the Tonks limit: $k^{2}\mathcal{E}_k(r)$
vs. $r/k$ at fixed $\rho=0.75$ and $a=400$ shows data collapse for large
$k$ according to Eq. (\ref{erscalingform}).}
\label{erscalingfig}
\end{center}
\end{figure}
\begin{figure}
\begin{center}
\includegraphics[width=80mm]{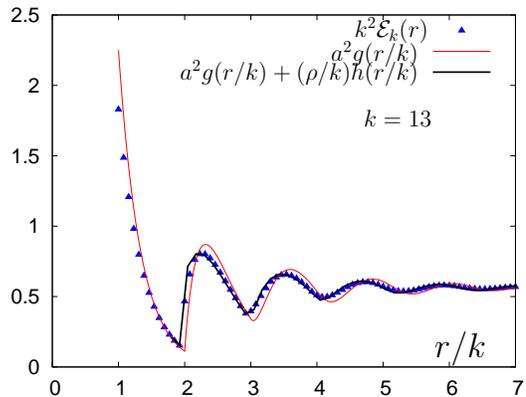}
\caption{(Color online) Comparison of $k^2\mathcal{E}_k(r)$ with the
Tonks result, $a^2g(r/k)$, for $k=13$, $a=400$, and $\rho=0.75$, showing
the discrepancy between the two for large but finite $k$. The
discrepancy is resolved on including the correction to the Tonks result,
to leading order in $1/k$, as is shown by a comparison of
$k^2\mathcal{E}_k(r)$ with the function $a^2g(r/k)+(\rho/k)h(r/k)$ (Eq.
(\ref{Tonkscorrection})).}
\label{erscalingfigzoom}
\end{center}
\end{figure} 

A quantity of physical interest for the equilibrium Tonks gas is the two-particle distribution $P(R_1,R_2)$, defined such 
that $P(R_1,R_2)dR_1dR_2$ is the joint probability of finding the rightmost end of one hard rod 
between $R_1$ and $R_1+dR_1$ and that of another between $R_2$ and $R_2+dR_2$. For a translationally invariant system, $P(R_1,R_2)$ is a function of the separation 
$R\equiv|R_2-R_1|$. Then, if $a$ is the rod length, it is known that \cite{Salsburg:1953}
\bea
P(R)&=&g(x), \nonumber \\
g(x)&\equiv&\frac{1}{la^2}\sum_{m=1}^\infty A(x-m) \frac{(x-m)^{m-1}}{(m-1)!(l-1)^m}\nonumber \\
&&\times\exp\Big(-\frac{x-m}{l-1}\Big), 
\label{Tonkstwopt}
\eea      
where $x=R/a$ is a reduced distance. Here, $l=1/a \rho_\mathrm{T}$, where $\rho_\mathrm{T}$ is the 
density of rods, and $A(x)$ is the unit step function:
\bea
A(x)&=&\left\{ 
\begin{array}{ll}
               0 \mbox{~for $x < 0$}, \\
               1 \mbox{~for $x \ge 0$}. \\
               \end{array}
        \right. 
\label{unitstep} 
\eea

We now show that the continuum limit of the $k$-ASEP engine-engine
correlation easily yields Eq. (\ref{Tonkstwopt}). Such a derivation is
justified by the fact that, as discussed in Sec. \ref{steadystate}, the
$k$-ASEP has the same steady-state measure of configurations for both unbiased and biased motion of the k-mers.
This fact further implies that Eq. (\ref{Tonkstwopt}) also holds for the driven Tonks gas.

The continuum limit of the $k$-ASEP is obtained by considering the joint limit $k \rightarrow \infty$, $r \rightarrow \infty$, 
and the lattice spacing $\eps \rightarrow 0$, while keeping $R=r\eps$, $a=k\eps$, and $\rho$ fixed and finite \cite{Schonherr:2005}. The $k$-ASEP then 
describes biased motion of hard rods of length $a$ along a continuous line. The density of hard rods is $\rho_{\mathrm{T}}=\rho/a$. 
In this limit, when $\rho_0 = \rho\eps/[a(1-\rho)]$ and $(1-\rho_0) = \exp\Big[-\rho\eps/[a(1-\rho)]\Big]$, Eq. (\ref{erthermodynamiclimit}) reduces to
\bea
\mathcal{E}_k(r) &=&  
\frac{\eps^2}{la^2}\sum_{m=0}^{\lfloor {R/a-1} \rfloor}\frac{[R/a-(m+1)]^m}{m! (l-1)^{m+1}} \nonumber \\
&&\times \exp\Big(-\frac{R/a-(m+1)}{l-1}\Big). 
\label{erscalinglimit}
\eea
Comparing the right-hand side of the last equation with Eq. (\ref{Tonkstwopt}), and noting that $R/a=r/k$, we find that
\be
\mathcal{E}_k(r) = \delta^2g\Big(\frac{r}{k}\Big).
\label{erscalinglimit1}
\ee
Now, since $\delta=a/k$, we find that in the continuum limit, i.e., in the limit $k \rightarrow \infty$, $r \rightarrow \infty$, $\eps \rightarrow 0$, while  keeping $a$, $R$, and $\rho$ fixed and finite, $\mathcal{E}_k(r)$ for different $k$ has the scaling form
\be
\mathcal{E}_k(r) = \frac{a^2}{k^2}g\Big(\frac{r}{k}\Big).
\label{erscalingform}
\ee
Moreover, in the continuum limit, defining $P(R)=\mathcal{E}_k(r)/\eps^2$, we find 
from Eq. (\ref{erscalinglimit1}) that $P(R)$ is precisely in the form of Eq. (\ref{Tonkstwopt}). We have thus derived the 
two-particle distribution, valid for both the equilibrium and the driven Tonks gas, by considering the continuum limit of the $k$-ASEP. 

Figure \ref{erscalingfig} shows plots of  $k^2 \mathcal{E}_k(r)$ for
different $k$ at fixed $\rho$ and $a$, evaluated using Eq.
(\ref{erthermodynamiclimit}). As $k$ increases, the curves show a good
scaling collapse, in accordance with Eq. (\ref{erscalingform}). For
large but finite $k$, the right-hand side of Eq. (\ref{erscalingform})
has finite-$k$ corrections, as is suggested by the discrepancy between
$k^2\mathcal{E}_k(r)$ and $a^2g(r/k)$, shown in Fig.
\ref{erscalingfigzoom}. The leading-order correction to Eq. (\ref{erscalingform}) will be discussed in the following subsection.  

The behavior of the density correlation $C_k(r)$ in the continuum limit may be easily obtained by using 
Eqs. (\ref{crcompact}) and (\ref{erscalingform}). We find that in this limit, for different $k$ at fixed $\rho$ and $a$, we have
\be
C_k(r) = \frac{a^2(1-\rho)^2}{\rho^2}g\Big(\frac{r}{k}+1\Big)+(2\rho-1).
\label{crscalingform}
\ee

\begin{figure}
\begin{center}
\includegraphics[width=80mm]{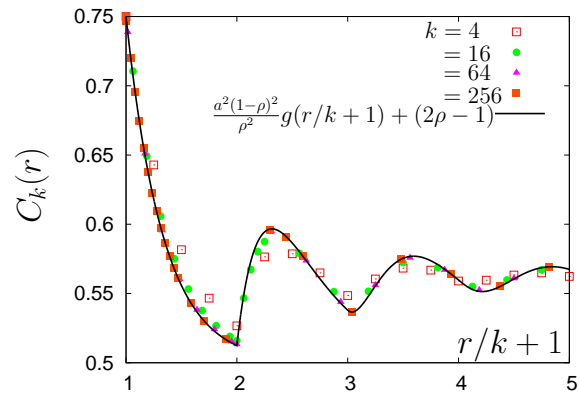}
\caption{(Color online) Scaling approach of the density-density correlation $C_k(r)$ to the Tonks limit: $C_k(r)$ vs. $r/k+1$ at fixed $\rho=0.75$ and $a=400$ exhibits data collapse for large $k$ in accordance with Eq. (\ref{crscalingform}). The data are obtained by numerically evaluating Eqs. (\ref{erthermodynamiclimit}) and (\ref{crexpression}).}
\label{crscalingfig}
\end{center}
\end{figure} 

Figure \ref{crscalingfig} shows plots of $C_k(r)$ for different $k$ at fixed $\rho$ and $a$, evaluated using Eqs. (\ref{erthermodynamiclimit}) and (\ref{crexpression}). As $k$ increases, the curves show a good scaling collapse, in accordance with Eq. (\ref{crscalingform}). 
\subsection{Finite-$k$ corrections to Tonks two-particle distribution}
In order to compute finite-$k$ corrections to Eq. (\ref{erscalingform}),
we evaluate the engine-engine correlation for finite $r$, $k$, and
$\eps$, with $r \gg 1$, $k \gg 1$, and $\eps \ll 1$, keeping $R=r\eps,
a=k\eps$, and $\rho$ fixed and finite. Then, on substituting $\rho_0 \approx \fr{\rho}{k(1-\rho)}-\fr{\rho^2}{k^2(1-\rho)^2}$ and $(1-\rho_0) \approx \exp\Big(-\fr{\rho}{k(1-\rho)}\Big)\exp\Big(\fr{\rho^2}{2k^2(1-\rho)^2}\Big)$ into Eq. (\ref{erthermodynamiclimit}), and keeping terms to leading order in $1/k$, we get
\be
k^2\mathcal{E}_k(r) \approx a^2 g\Big(\fr{r}{k}\Big)+\fr{\rho }{k}h\Big(\fr{r}{k}\Big),
\label{Tonkscorrection}
\ee
where 
\bea
h(x)&=&\sum_{m=1}^\infty A(x-m)\fr{(x-m)^{m+1}}{(m-1)!(l-1)^m}\exp\Big(-\fr{x-m}{l-1}\Big) \nonumber \\
&&\times \Big[\fr{ m(m-1)}{2(x-m)}-\fr{m}{(l-1)}+\fr{x-m}{2(l-1)^2}\Big]; ~~~~x \ne 1. \nonumber \\
\label{g}
\eea
Here, $A(x)$ is the unit step function defined in Eq. (\ref{unitstep}).
Figure \ref{erscalingfigzoom} shows that inclusion of the leading-order correction to
the Tonks result, as in Eq. (\ref{Tonkscorrection}), indeed resolves the discrepancy between $k^2\mathcal{E}_k(r)$ and $a^2g(r/k)$.
\section{Steady state dynamics}
\label{steadystatedynamics}
\subsection{Current and kinematic wave velocity}
\label{currentvK}
In discussing the current in the system, we need to distinguish between
that associated with the motion of engines, and that with the $k$-mers.
Contributions to the engine current across a bond $(i,i+1)$ arise when
either (i) the $i$-th site is occupied by an engine, while the
$(i+1)$-th site is vacant, or, (ii) the $(i+1)$-th site is occupied by
an engine, while the $(i-k+1)$-th site is vacant. On using the results
of Sec. \ref{steadystate}, we find that in the thermodynamic limit, the average engine current in the steady state is given by \cite{Shaw:2003}
\be
J_e=\frac{(p-q)\rho(1-\rho)}{\rho+k(1-\rho)}.
\label{J_e}
\ee  

To compute the $k$-mer current, $J$, note that associated with the motion of the engine to an adjacent site is the sliding of the corresponding $k$-mer across $(k-1)$ bonds, so that $J=kJ_e$. It can be checked that $J$ has a maximum at the density $\rho_{\rm c}=\sqrt{k}/(\sqrt{k}+1)$. 

The kinematic wave velocity $v_{K}\equiv \partial J/\partial \rho$ accounts for the transport of density 
fluctuations through the system in the steady state \cite{Lighthill:1955}. We find 
\be
v_{K}=k(p-q)\left[\frac{(k-1)\rho^2+k(1-2\rho)}{[\rho+k(1-\rho)]^{2}}\right].
\label{vk}
\ee
Evidently, $v_K$ vanishes if the density is $\rho_{\rm c}$. \rm As we discuss below, $v_K$ plays an important role in determining the form of the temporal decay of the density autocorrelation.
\subsection{Mapping to the ASEP: Wheeling velocity}
\label{mappingtoASEP}
\begin{figure}[h]
\begin{center}
\includegraphics[width=90mm]{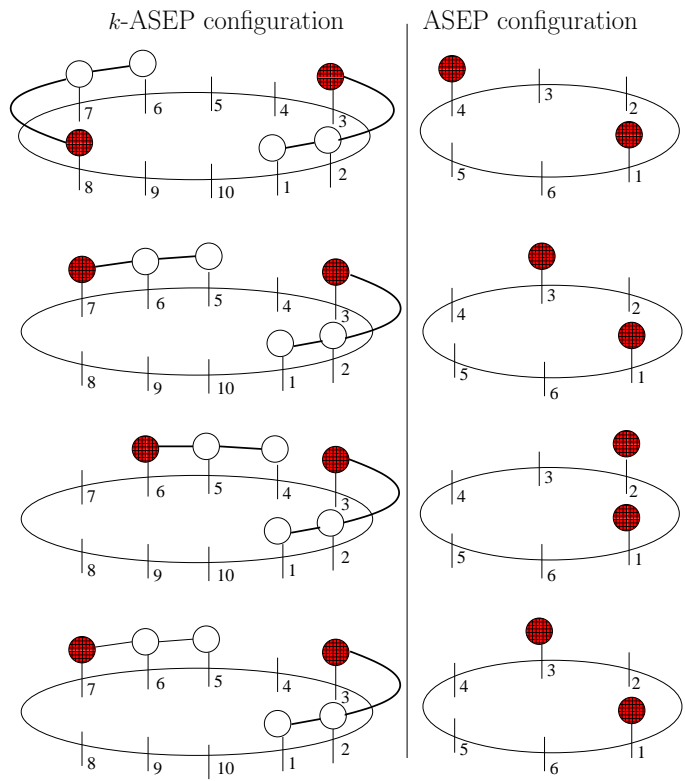}
\end{center}
\caption{(Color online) Mapping of a typical $k$-mer configuration to an ASEP configuration and 
their subsequent evolution in time. The first site of the ASEP lattice may be defined in more than one way; 
the figure illustrates one possibility.}
\label{kmerasepmapping}
\end{figure}
We now discuss a mapping of the $k$-ASEP with a number, $N$, of $k$-mers
on a 1D periodic lattice of $L$ sites to an ASEP of $N$ hard-core
particles on a 1D periodic lattice of $L'=L-N(k-1)$ sites. We show that
as a result of the mapping, a fixed site in the $k$-ASEP corresponds to
an ASEP site that moves around the ASEP ring with a finite mean velocity. This phenomenon is known as wheeling, and the mean velocity is called the wheeling velocity $W$ \cite{Barma:2007}. This velocity plays an important role in the scaling properties of the temporal density-density correlation of the $k$-ASEP, as we discuss in the next subsection. 

The mapping involves representing each $k$-mer by a hard-core particle that corresponds to the engine of the $k$-mer, as illustrated in Fig. \ref{kmerasepmapping}. In this way, every $k$-ASEP configuration is mapped to a unique configuration in the ASEP.  Associated with the motion of a $k$-mer is that of the corresponding ASEP particle according to the ASEP dynamics. We now see that the quantity $\rho_0$ in Eq. (\ref{eq:rho0}) is the particle density in the equivalent ASEP.

It can be seen from Fig. \ref{kmerasepmapping} that in the process of mapping, the ASEP 
image of a fixed $k$-mer site moves around the ASEP lattice as a result of the $k$-mer motion. For example, 
consider the transition $(111\ldots1)0 \rightarrow 0(111\ldots1)$. It is easy to see
 that in this transition, the ASEP images of $k$-ASEP sites containing the $0$ and the leftmost $1$ in the string $(111\ldots1)0$ 
do not change, while the images of those containing the remaining $1$'s in the string increase by one unit. On the other hand, 
in the transition $0(111\ldots1) \rightarrow (111\ldots1)0$, the ASEP images of $k$-ASEP sites containing the $0$ 
and the rightmost $1$ in the string $0(111\ldots1)$ do not change, while the images of those containing the remaining $1$'s in the string 
decrease by one unit. This motion of an ASEP site corresponding to a fixed site in the $k$-ASEP is the phenomenon of wheeling. As a result, the displacement of the ASEP image of a fixed $k$-ASEP site in time $t$ is given by
\be
\Delta r(t)=Wt+\phi(t),
\ee
where $\phi(t)$ is a random variable with zero mean, arising from the
stochasticity in the dynamics. Referring to the results on joint
occupation probabilities in Sec. \ref{steadystate}, we find that the wheeling velocity is given by
\be
W=\frac{(p-q)(k-1)\rho(1-\rho)}{\rho+k(1-\rho)}.
\label{wheelingvelocity}
\ee 
  
\subsection{The temporal density-density correlation}
\begin{figure}[h!]
\begin{center}
\includegraphics[width=90mm]{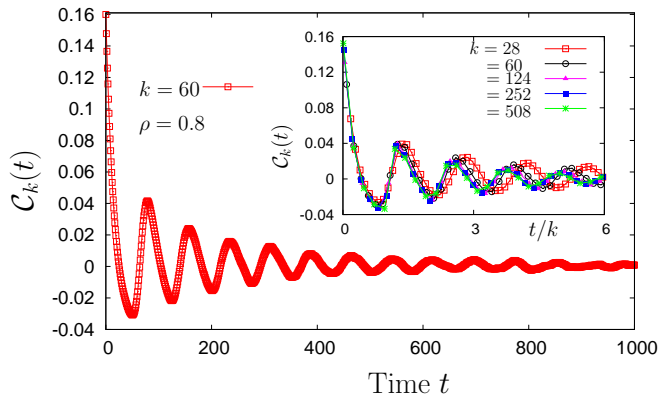}
\caption{(Color online) Short-time behavior of the $k$-ASEP density autocorrelation $\mathcal{C}_k(t)$ for $k=60$ and $\rho=0.8$. The data are obtained from Monte Carlo simulations. The inset shows $\mathcal{C}_k(t)$ vs. $t/k$ at fixed $\rho=0.8$, illustrating scaling for large $k$.}
\label{ctscalingfig}
\end{center}
\end{figure}

The temporal density-density correlation function in the steady state of the $k$-ASEP is defined as
\be
\mathcal{C}_k(r=|i-j|,t)\equiv \langle n_{i}(0)n_{j}(t) \rangle -
\rho^2,
\label{ct}
\ee  
where $n_{i}(t)$ denotes the occupation index of site $i$ at time $t$. In particular, the density autocorrelation is given by $\mathcal{C}_k(t) \equiv \mathcal{C}_k(0,t)$. 

We study $\mathcal{C}_k(t)$ by performing Monte Carlo simulations of the
$k$-ASEP in the steady state. Figure \ref{ctscalingfig} shows that
$\mathcal{C}_k(t)$ oscillates in time. The inset shows that for large $k$,
the autocorrelation for different $k$ at a fixed density $\rho$ is
initially a function of $t/k$. This behavior holds up to a $k$-dependent
time. To understand the dependence of $\mathcal{C}_k(t)$ on the ratio $t/k$, we note that at short times, the relevant time scale 
is set by the time $\tau(k)$ that an occupied site takes to fall vacant. This is
consistent with $\mathcal{C}_k(t)$ being a function of $t/\tau(k)$.
An upper bound on $\tau(k)$ may be estimated as the time $\tau_e(k)
\approx k/v_e$, the time that an engine takes to move by its own length.
Here $v_e$ is the velocity of an engine, which may be obtained from Eq.
(\ref{J_e}) as $v_e=\frac{k(p-q)(1-\rho)}{\rho+k(1-\rho)}$. In the limit
of large $k$, one finds that $\tau_e(k) \approx k$, so that
$\mathcal{C}_k(t)$ is a function of $t/k$, as observed.     

We now discuss the behavior of $\mathcal{C}_k(t)$ at long times. To
proceed, we examine the function $\mathcal{C}_k(r,t)$ for which an
earlier study in the case $k=2$ has illustrated that in the limit of
long times and large distances, it assumes a particular scaling form
\cite{Barma:2007}. To obtain the scaling for general $k$, we utilize the
mapping to the ASEP discussed in Sec. \ref{mappingtoASEP} and invoke known scaling properties of the density correlation in the latter.

\begin{figure}[h!]
\begin{center}
\includegraphics[width=80mm]{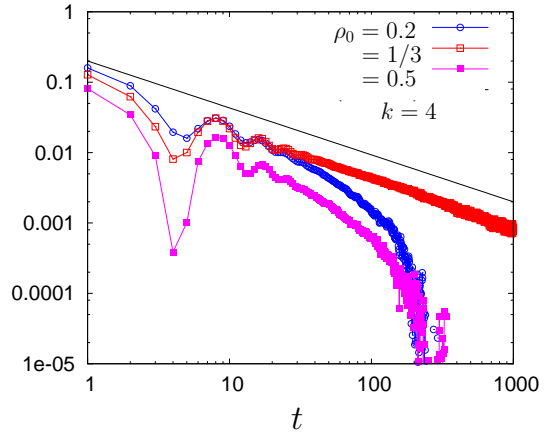}
\caption{(Color online) Long-time decay of the $k$-ASEP density autocorrelation $\mathcal{C}_k(t)$, as a power law in time at the compensating density $\rho_{0c}$, and as an exponential at other densities. The data are obtained from Monte Carlo simulations. Here $k=4$, so that $\rho_{0c}=1/3$. The black line has the slope of $-2/3$.}
\label{ctfig}
\end{center}
\end{figure} 

In the ASEP ($k=1$), the temporal density-density correlation function, $\mathcal{C}_1(r=|i-j|,t)\equiv \langle n_{i}(0)n_{j}(t) \rangle - \rho_0^{2}$, 
in the scaling limit follows the form \cite{Spohn:2002}
\be
\mathcal{C}_1(r,t) \propto t^{-2/3} F(u); ~~~~~~u=\frac{1}{2}(J_0t^{2})^{-1/3}(r-v_{0K}t).
\label{C0rt}
\ee  
Here, $J_0$ and $v_{0K}$ are, respectively, the steady-state current and the kinematic wave velocity in the ASEP, given by $J_0=(p-q)\rho_0(1-\rho_0)$, and $v_{0K}=(p-q)(1-2\rho_0)$ \cite{Schutz:2001}. 
In the limit of large $u$, it is known that $F(u) \sim
\exp(-\mu|u|^{3})$, with $\mu \simeq -0.295$ \cite{Spohn:2002}. 

It is evident from the $k$-ASEP to ASEP mapping discussed above that at long times, neglecting the stochastic part $\phi(t)$ in the displacement of a mapped site in the ASEP, the correlation $\mathcal{C}_k(r,t)$ has a behavior similar to $\mathcal{C}_1(r+Wt,t)$ \cite{Barma:2007}. Thus, for a fixed $k$, in the limit of long times and large distances, $\mathcal{C}_k(r,t)$ follows the scaling form
\be
\mathcal{C}_k(r,t) \propto t^{-2/3} F(u'); ~~~~~~u'=\frac{1}{2}(J_0t^{2})^{-1/3}\Big(r+(W-v_{0K})t\Big).
\ee 
The autocorrelation $\mathcal{C}_k(t)$ behaves asymptotically as 
\be
\mathcal{C}_k(t) \propto t^{-2/3}e^{-\kappa t},
\label{Ctlongtimes}
\ee
where $\kappa$ is a constant determined by the difference $(W-v_{0K})$. Thus, at long times, $\mathcal{C}_k(t)$ decays as an exponential in time, unless the density $\rho$ is such that the difference vanishes. In this case, the autocorrelation at late times decays in time as a power law: $\mathcal{C}_k(t) \sim t^{-2/3}$. The corresponding ASEP density is called the 
compensating density $\rho_{0c}$ \cite{Barma:2007}, and satisfies
\be
\rho_{0c}^{2}(k-1)+2\rho_{0c}-1=0.
\ee
Solving for the positive root, we get
\be
\rho_{0c}=\frac{1}{\sqrt{k}+1},
\ee
which matches with the result for $k=2$ derived in \cite{Barma:2007}.
Note that corresponding to $\rho_{\rm 0c}$ is the $k$-ASEP density of
occupied sites $\rho_{c}$, mentioned in Sec. \ref{currentvK}, at which the $k$-mer current $J$ is maximized and the kinematic wave velocity $v_{K}$ is zero.

Figure \ref{ctfig} shows $\mathcal{C}_k(t)$ as a function of time for three 
values of the ASEP density, namely, the compensating density $\rho_{0c}$, and two other
 values on either side. We see an asymptotic $t^{-2/3}$ decay 
of the autocorrelation at the compensating density and an exponential decay at other densities, in accordance with our analysis above. 

\section{Acknowledgement}
Part of this work is based on the Ph. D. thesis of SG at the Tata Institute of Fundamental Research, Mumbai. He acknowledges support of the Israel Science Foundation (ISF), and the French contract ANR-10-CEXC-010-01. We thank G. M. Sch\"{u}tz and R. K. P. Zia for discussions and for pointing out Ref. \cite{Schonherr:2005} and Ref. \cite{Dongthesis}, respectively, to us.
\appendix
\section{Mapping of the $k$-ASEP to the ZRP}
\label{ZRPmapping}
In this appendix, we discuss a mapping of the $k$-ASEP to a zero-range
process (ZRP) and show how static correlations in the former are
obtained by using the mapping. This method of obtaining the correlations
is an alternative to the one discussed in Sec. \ref{steadystatestatics}. 

 \begin{figure}
\begin{center}
\includegraphics[width=80mm]{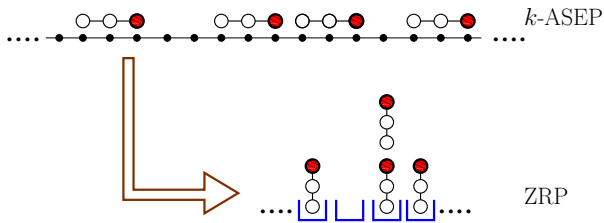}
\caption{(Color online) Mapping of the $k$-ASEP to the ZRP.  The $k$-ASEP vacancies are considered as sites in the ZRP, while an uninterrupted sequence of $k$-mers  
in front of a vacancy is regarded as a set of particles occupying the corresponding ZRP site, such that the number of particles equals the number of
$k$-mers in the sequence.}
\label{fig:zrp}
\end{center}
\end{figure} 

In the ZRP, an unrestricted number of particles resides on lattice sites
and hops between sites with a rate that depends only on the number of particles on the departure site \cite{Evans:2005}.
Each $k$-ASEP configuration can be mapped to a unique ZRP configuration in the following way: one considers vacancies ($0$'s) in the $k$-ASEP as sites in the ZRP and an uninterrupted sequence of $k$-mers following a vacancy as particles residing on the corresponding ZRP site, with the number of particles equal to the number of $k$-mers in the sequence. The ZRP has $N$ particles and $M \equiv L-Nk$ sites labeled by the index $i$. A generic ZRP configuration is the set $\{m_i\} \equiv (m_1,m_2 \dots m_M)$, where $m_i$ is the number of particles on the $i$-th site (see Fig. \ref{fig:zrp}). The motion of a $k$-mer in the $k$-ASEP translates to hopping of a particle from a ZRP site to its right or left neighbor with rates $p$ or $q$, respectively. 

For an arbitrary hop rate, the ZRP has a product measure stationary
state \cite{Evans:2005}. In our case, where the rates do not depend on
the number of particles on the departure site, the steady-state weight of any configuration $\{m_i\}$ is
\bea
P(\{m_i\}) = \prod_i f(m_i)\delta\Big(N-\sum\limits_{i=1}^M m_i\Big), ~~f(m) = 1.
\label{eq:ED}
\eea
The delta function stands for overall particle conservation.

The steady state weight of any $k$-ASEP configuration is obtained by
mapping it to a configuration in the equivalent ZRP and computing its
weight by utilizing Eq. (\ref{eq:ED}). In our case, since  $f(m)=1$ for
all $m$, all $k$-ASEP configurations with a given number, $N$, of
$k$-mers are equally likely. Now, using the formalism discussed in
\cite{Basu:2010}, one may rewrite the steady-state weight of any $k$-ASEP configuration $(n_1,n_2,\ldots,n_L)$ in a matrix product form by replacing each occupation number $n_i$ by either a matrix $D$ or a matrix $E$ depending on whether $n_i$ is $1$ or $0$, respectively. From the correspondence between the $k$-ASEP and the ZRP, we have
\bea
\hspace{-0.8cm}P(\{m_i\})&=&{\rm Tr}[ED^{km_1}\ldots ED^{km_M}]\delta\Big(N-\sum\limits_{i=1}^M m_i\Big),
\label{eq:eqv}
\eea
where ${\rm Tr}$ denotes the usual matrix trace operation.

Without loss of generality, one may take $E = \ket\alpha\bra\beta$, where the vectors $\ket\alpha$ and $\bra\beta$ are to be determined. This choice of $E$ together with Eqs. (\ref{eq:eqv}) and (\ref{eq:ED}) demand that for any positive integer $m$, the matrix $D$ satisfies
\bea
 \bra \beta  D^{mk} \ket \alpha=f(m)=1. 
\label{eq:main} 
\eea
Also, $\bra \beta  D^{j} \ket \alpha = 0$ for positive integers $j$ which are not multiples of $k.$ A simple $k$-dimensional representation of matrices $E$ and $D$ is when they have non-zero elements $E_{1 1}=1$ and $D_{k 1}=1=D_{i, i+1}$, that is,
\bea
&&E= \ket \alpha \bra \beta, \; \; {\rm  with} ~~~~ \ket\alpha=\ket 1, ~~\bra\beta=\bra 1, \nonumber \\ 
&&D= \sum_{i=1}^{k-1} \ket i \bra {i+1} + \ket k \bra 1.
\eea
Here, the set $\{ \ket i\}$ represents the standard basis vectors in $k$-dimensions.
This choice ensures that the weight of any configuration with one or more blocks of $l$ particles is zero if $l$ is not an integral multiple of $k$; all other configurations are equally probable.

Let us mention that the matrix formulation discussed above is different from the Matrix Product Ansatz (MPA) of Derrida {\it et al.} \cite{MPA}, in which matrices satisfy specific algebraic relations dictated by the system dynamics.  For models with ZRP correspondence, the matrices generically satisfy Eq. (\ref{eq:main}), and therefore, depend only on the weights $f(m).$ It is always possible to get one representation of these matrices, whereas finding explicit representation of the MPA matrices is non-trivial.
  
{\it Partition function  $Z_L(z)$:} The first task in computing $k$-ASEP static correlations is to find the partition function of the system, which is conveniently done in the grand canonical ensemble by associating the fugacity $z$ with any occurrence of the matrix $D$. This gives $Z_L(z) ={\rm Tr}[C^L]$, where $C=zD +E$. The configuration with no vacant site is not dynamically accessible. Thus, $Z_L(z)$ is given by weights of all configurations with at least one vacant site:
\bea
Z_L(z) &=& \sum_{n=1}^L {\rm Tr}\left[(zD)^{n-1}EC^{L-n}\right] \nonumber \\
&=&\sum_{n=1}^L\bra \beta C^{L-n} (zD)^ {n-1} \ket \alpha.
\label{eq:ZL}
\eea

To proceed, we use the following generating function:  
\bea
\mathcal{Z}(z,\gamma) &=&  \sum_{L=1}^\infty \gamma^L Z_L(z)
= \bra \beta\frac{\gamma}{{\cal I}-\gamma C}  \frac{1}{{\cal I}-\gamma z D}  \ket \alpha \cr
& =& {\gamma [1 + (k-1) (\gamma z)^k] \over [1 - \gamma - (\gamma z)^k][1
-(\gamma z)^k]},
\label{eq:Zzg}
\eea
where ${\cal I}$ is the $k$-dimensional identity matrix. Note that $\mathcal{Z}(z,\gamma)$ may be interpreted as the partition function in the variable length ensemble. The parameters $z$ and $\gamma$ together determine macroscopic observables like the density of occupied sites and the average system size. The density of occupied sites is $\rho= 1- \la \bar n_i\ra$, where
\bea
\la \bar n_i\ra   =  {\gamma \over \mathcal{Z}}\bra \beta\frac{1}{{\cal I}-\gamma C}\ket \alpha  
 = {1 - (\gamma z)^k \over 1 + (k-1) (\gamma z)^k}.\label{eq:G1}
\eea
The average system size is given by $\la L\ra = \frac{\gamma}{\mathcal{Z}} \frac{\partial \mathcal{Z}}{ \partial \gamma }.$

One may check that $\la L\ra$ has the radius of convergence   
\bea
z^*= \frac 1  \gamma (1-\gamma )^{1/k}.
\label{eq:z*}
\eea
Thus, the thermodynamic limit $\la L\ra \to \infty$ is achieved  at $z=z^*$, when every observable of the system becomes a function of $\gamma$ {\it only}.  
For example, $\rho$ is obtained from Eq. \eqref{eq:G1} as $\et =\frac{k(1-\gamma)}{\gamma + k(1-\gamma)}$, which may be inverted to obtain
 \bea
 \gamma =\frac{k(1-\et)}{\et + k(1-\et)} = 1-\rho_0.\label{eq:gamma} 
\eea

{\it Equal-time correlations $\mathit{V}_k(r),\mathcal{E}_k(r)$:} One may compute $\mathit{V}_k(r)$ by writing it in terms of matrices as follows:  
\bea
\mathit{V}_k(r)  & =& {\gamma^{r+1} \over \mathcal{Z}} \bra \beta   C^{r-1} 
\ket \alpha \bra  \beta\frac{1}{{\cal I}-\gamma C}\ket \alpha \cr 
&=& \gamma^r (1-\et) \bra \beta C^{r-1} \ket  \alpha.
\label{eq:corr}
\eea
Now, for any integer $j$, one may check that
\bea
\bra \beta  C^j \ket \alpha &=& \sum_{m=0}^{\lfloor{j/k}\rfloor}\left({j - mk + m \atop m}\right){z}^{mk}, \label{eq:cj}
\eea
which results in
\bea
\mathit{V}_k(r) = \gamma^r (1-\et) \sum_{m=0}^{\lfloor{(r-1)/k}\rfloor}\Omega^{\mathrm{free}}_{m,r-1}{z}^{mk}. \nonumber
\eea 

The equal-time engine-engine correlation ${\cal E}_k(r)$ may be similarly calculated by using the matrix formulation: 
\bea
{\cal E}_k(r)&  =&  {\gamma^{r+k}\over \mathcal{Z}} \bra  \beta\frac{1}{{\cal I}-\gamma C}
\ket \alpha \sum_{m=0}^{\lfloor{(r-k)/k}\rfloor}\Omega^{\mathrm{free}}_{m,r-k}z^{(m+2)k}\cr
 & = & (1 -\rho) \gamma^{r+k-1} \sum_{m=0}^{\lfloor{(r-k)/k}\rfloor}
 \Omega^{\mathrm{free}}_{m,r-k}z^{(m+2)k}. 
\eea

In the thermodynamic limit, using Eq. (\ref{eq:z*}), we get 
\bea
\mathit{V}_k(r)=\frac{\gamma^{r+1}}{\gamma+ k(1-\gamma)}
\sum_{m=0}^{\lfloor {(r-1)/k}\rfloor}\Omega^{\mathrm{free}}_{m,r-1} \left({1 -\gamma \over
\gamma^k}\right)^m,&& \nonumber \\
{\cal E}_k(r)={\gamma^{r+k} (1-\gamma)^2\over\gamma+ k(1-\gamma)
}\sum_{m=0}^{\lfloor{(r-k)/k}\rfloor}
\Omega^{\mathrm{free}}_{m,r-k}\left({1 -\gamma \over \gamma^k}\right)^m.&& \nonumber
\eea
Replacing $\gamma$ by $1-\rho_0$ in the expressions on the right reduce them to those in Eqs. \eqref{vrthermodynamiclimit}
and \eqref{erthermodynamiclimit}, respectively.

\end{document}